\documentclass[journal,twoside,web]{ieeecolor}
\usepackage{generic}
\usepackage{cite}
\usepackage{amsmath,amssymb,amsfonts}
\usepackage{algorithmic}

\usepackage{graphicx}
\usepackage{multirow}

\usepackage{hyperref}

\usepackage{subcaption}
\usepackage{textcomp}
\usepackage[ruled,vlined, linesnumbered]{algorithm2e}

\def\BibTeX{{\rm B\kern-.05em{\sc i\kern-.025em b}\kern-.08em
    T\kern-.1667em\lower.7ex\hbox{E}\kern-.125emX}}
\begin{document}
\title{Fast-DDPM: Fast Denoising Diffusion Probabilistic Models for Medical Image-to-Image Generation}
\author{Hongxu Jiang, Muhammad Imran, Teng Zhang, Yuyin Zhou, Muxuan Liang, Kuang Gong, and Wei Shao
\thanks{This work was supported by the Department of Medicine and the Intelligent Clinical Care Center at the University of Florida College of Medicine. The authors express their sincere gratitude to the NVIDIA AI Technology Center at the University of Florida for their invaluable feedback, technical guidance, and support throughout this project. (Corresponding author: Wei Shao.)}
\thanks{Hongxu Jiang and Teng Zhang are with Department of Electrical and Computer Engineering at University of Florida, Gainesville, FL 32610, USA (e-mail: hongxu.jiang@ufl.edu; teng.zhang@ufl.edu). }
\thanks{Muhammad Imran is with Department of Medicine at University of Florida, Gainesville, FL 32610, USA (e-mail: muhammad.imran@ufl.edu).}
\thanks{Yuyin Zhou is with Department of Computer Science and Engineering at University of California, Santa Cruz, CA 95064, USA (e-mail: yzhou284@ucsc.edu).}
\thanks{Muxuan Liang is with Department of Biostatistics at University of Florida, Gainesville, FL 32610, USA (e-mail: muxuan.liang@ufl.edu).}
\thanks{Kuang Gong is with Department of Biomedical Engineering at University of Florida, Gainesville, FL 32610, USA (e-mail: kgong@bme.ufl.edu).}
\thanks{Wei Shao is with Department of Electrical and Computer Engineering, Department of Medicine, and Intelligent Clinical Care Center at University of Florida, Gainesville, FL 32610, USA (e-mail: weishao@ufl.edu).}
}

\maketitle

\begin{abstract}
Denoising diffusion probabilistic models (DDPMs) have achieved unprecedented success in computer vision. However, they remain underutilized in medical imaging, a field crucial for disease diagnosis and treatment planning. This is primarily due to the high computational cost associated with the use of large number of time steps (e.g., 1,000) in diffusion processes. Training a diffusion model on medical images typically takes days to weeks, while sampling each image volume takes minutes to hours. To address this challenge, we introduce Fast-DDPM, a simple yet effective approach capable of improving training speed, sampling speed, and generation quality simultaneously. Unlike DDPM, which trains the image denoiser across 1,000 time steps, Fast-DDPM trains and samples using only 10 time steps. The key to our method lies in aligning the training and sampling procedures to optimize time-step utilization. Specifically, we introduced two efficient noise schedulers with 10 time steps: one with uniform time step sampling and another with non-uniform sampling. We evaluated Fast-DDPM across three medical image-to-image generation tasks: multi-image super-resolution, image denoising, and image-to-image translation. Fast-DDPM outperformed DDPM and current state-of-the-art methods based on convolutional networks and generative adversarial networks in all tasks. Additionally, Fast-DDPM reduced the training time to 0.2$\times$ and the sampling time to 0.01$\times$ compared to DDPM. Our code is publicly available at: \url{https://github.com/mirthAI/Fast-DDPM}.
\end{abstract}

\begin{IEEEkeywords}
Conditional diffusion models, Fast-DDPM, Image-to-image generation, Deep learning
\end{IEEEkeywords}

\section{Introduction}
\label{sec:introduction}

Diffusion models~\cite{ho2020denoising, kingma2021variational, song2020denoising, song2020score} have become powerful tools for high-quality image generation~\cite{dhariwal2021diffusion}. The forward diffusion process incrementally adds noise to a high-quality image until it becomes pure Gaussian noise. An image denoiser is then trained to learn the reverse process, progressively removing noise from random Gaussian noise until it reconstructs a noise-free image. This enables the generation of new high-quality images that match the distribution of the training dataset. Beyond computer vision, diffusion models are increasingly applied to medical imaging~\cite{10.1007/978-3-031-43999-5_42,ozbey2023unsupervised,hu2023conditional,10.1007/978-3-031-19821-2_20} for improved disease diagnosis, treatment planning, and patient monitoring.

Training and sampling diffusion models are computationally expensive and time-consuming due to the large number of discrete time steps used to approximate the continuous diffusion process, as well as the need for large datasets to model complex image distributions. For 2D diffusion models, training can take days on a single GPU, limiting experimentation with different model architectures and hyperparameters. 
Sampling a diffusion model can take several minutes per 2D image, presenting challenges for real-time usability when applying 2D diffusion models to 3D image volumes consisting of hundreds of 2D slices, where the process may take several hours per volume.

Current research mainly focuses on accelerating the sampling process, using either training-free or training-based methods~\cite{zheng2023fast}. Training-free methods utilize efficient numerical solvers for stochastic or ordinary differential equations to reduce the number of time steps required for sampling. This approach can decrease the number of sampling time steps from 1,000 to 50-100, without compromising the quality of the generated images. Alternatively, training-based methods, such as progressive knowledge distillation~\cite{salimans2022progressive, meng2023distillation}, can further reduce the sampling steps to as few as 10. However, this approach requires the training of additional student models, which considerably increases the overall training cost and may not be practical for medical image analysis.

\begin{figure*}[!hbt]
\centering
\includegraphics[width=0.95\textwidth]{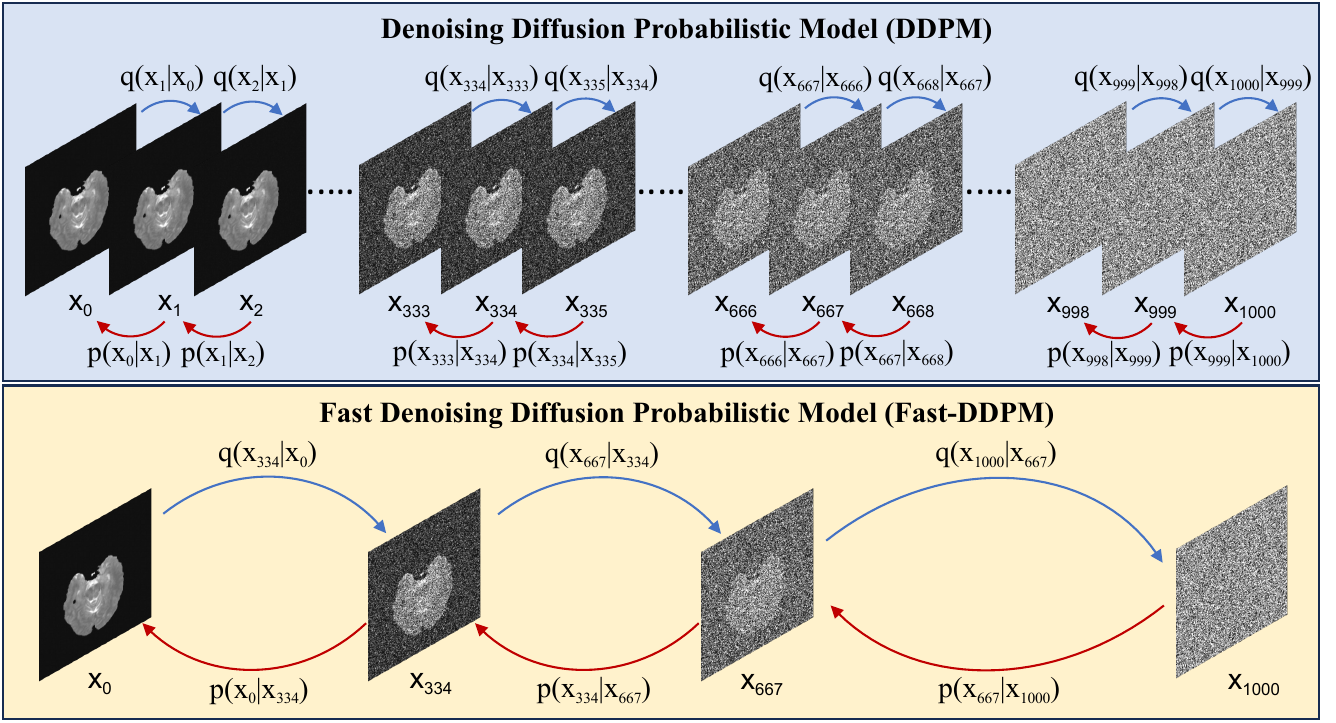}
\caption{Fast-DDPM uses significantly fewer time steps than DDPM. For illustration purposes, we used only 3 time steps for Fast-DDPM in this example.}
\label{fig:illustration}
\end{figure*}

In this paper, we propose Fast-DDPM, a model designed to reduce both training and sampling times by optimizing time-step utilization. We observed that the majority of time steps involved in training are not utilized by faster diffusion samplers during sampling. For instance, a Denoising Diffusion Probabilistic Model (DDPM) trained with 1,000 time steps is often sampled by the Denoising Diffusion Implicit Model (DDIM) using only 100 time steps. This practice is resource-inefficient, as the image denoiser is trained on 900 time points that are ultimately skipped during sampling. To address this inefficiency in training, we propose training and sampling the DDPM model at the same 10 time steps, as illustrated in Figure~\ref{fig:illustration}.

We evaluated Fast-DDPM on three medical image-to-image tasks, including multi-image super-resolution for prostate MRI, denoising low-dose CT scans, and image-to-image translation for brain MRI. Fast-DDPM achieved state-of-the-art performance across all three tasks. Notably, Fast-DDPM is approximately 100 times faster than the DDPM model during sampling and about 5 times faster during training. We summarize the major contributions of this paper as the following:
\begin{itemize}
\item Introduction of Fast-DDPM: A simple yet effective approach that simultaneously improves training speed, sampling speed, and generation quality.
\item Novel Training Strategy: Optimizes time-step utilization by aligning the training and sampling procedures, significantly enhancing both speed and accuracy.
\item Efficient Noise Schedulers: Designed two noise schedulers with 5-10 time steps—one utilizing uniform sampling and the other non-uniform sampling—tailored for specific tasks to enhance performance.
\end{itemize}

\section{Related Work}
\subsection{Fast Sampling of Diffusion Models}
Current methods to accelerate the sampling of diffusion models can be broadly classified into two categories: training-free and training-based. Training-free methods primarily focus on developing efficient diffusion solvers to solve reverse stochastic differential equations (SDEs) or their equivalent ordinary differential equations (ODEs). For instance, the Denoising Diffusion Implicit Models (DDIM) approach~\cite{song2020denoising} models the reverse diffusion process using an ODE and significantly reduces the number of required sampling steps to 50-100 without impacting the quality of the generated samples. Building on this, FastDPM~\cite{kong2021fast} establishes a bijective mapping between continuous diffusion steps and continuous noise levels and uses this mapping to construct an approximation of the reverse process between DDPM and DDIM with fewer steps. DPM-Solver~\cite{lu2022dpm} introduces a high-order solver for diffusion ODEs based on an exact solution formulation that simplifies to an exponentially weighted integral, achieving high-quality sample generation with just 10-50 sampling steps. The Differentiable Diffusion Sampler Search~\cite{watson2022learning} sampler uses the reparametrization trick and gradient rematerialization to optimize over a parametric family of fast samplers for diffusion models by differentiating through sample quality scores. In addition to these solver-based methods, quantization emerges as another promising training-free method for improved diffusion model sampling. PTQ4DM~\cite{Shang_2023_CVPR} introduces Post-Training Quantization (PTQ) with a novel calibration method named NDTC, aimed at reducing the computational cost of noise estimation. Concurrently, Q-Diffusion~\cite{Q-Diffusion_2023_ICCV} proposes timestep-aware calibration alongside split shortcut quantization, further refining the quantization process for efficiency gains. DeepCache~\cite{ma2024deepcache} avoids computing deep features repeatedly in the network by caching them, thereby accelerating the sampling process.

A common training-based approach utilizes knowledge distillation, wherein a teacher diffusion model is first trained with a large number of time steps, followed by the sequential training of multiple student models with progressively fewer time steps, mirroring the teacher model's behavior~\cite{salimans2022progressive, meng2023distillation}. This methodology reduces inference time steps from 1,000 to as few as 16 while maintaining sampling quality. Recently, adversarial diffusion distillation~\cite{sauer2025adversarial} has enabled real-time, single-step image synthesis by combining score distillation and adversarial loss, achieving state-of-the-art quality with just 1–4 sampling steps. Furthermore, the rectified flow technique achieves high-quality sampling in a single time step by straightening the trajectory of the reverse-time ODE~\cite{liu2022flow}, and consistency models~\cite{song2023consistency} produce superior sampling quality within a single step by directly mapping noise to an image, either through distillation from pre-trained diffusion models or as standalone generative models.

\subsection{Diffusion Models in Medical Imaging}
Diffusion models have been applied to various medical imaging tasks, including segmentation~\cite{hu2023conditional,kim2023diffusion,xing2023diffunet}, anomaly detection~\cite{10.1007/978-3-031-16452-1_4,10.1007/978-3-031-45673-2_37,xu2024maediff}, denoising~\cite{gong2023pet,9941138,10.1007/978-3-031-43898-1_1}, reconstruction~\cite{Liu_Reconstruction_2023_ICCV,GUNGOR2023102872,10385176}, registration~\cite{10.1007/978-3-031-19821-2_20}, super-resolution~\cite{9941138,10.1007/978-3-031-43898-1_1,10.1007/978-3-031-43999-5_42,WU2023104901}, and image-to-image translation~\cite{lyu2022conversion,ozbey2023unsupervised,Kim_2024_WACV}. Among these, the DDPM model is the most widely used in medical imaging due to its simplicity. Diffusion models can be categorized as 2D, 3D, or 4D based on input image dimensions. Currently, 2D models are the predominant choice in medical imaging due to their ease of implementation and lower memory requirements. 
However, when implementing diffusion models for 3D and 4D applications, training can take significantly longer, lasting anywhere from weeks to months. Improving the training and sampling processes is crucial for a broader adoption of diffusion models in medical imaging.

\subsection{Unconditional and Conditional Diffusion Models}
Diffusion models can be broadly categorized into unconditional and conditional models. Unconditional diffusion models generate data samples based solely on the learned data distribution, without incorporating external information or conditions~\cite{song2020denoising,ho2020denoising, nichol2021improved}. In contrast, conditional diffusion models generate data samples guided by additional input conditions, such as text, class labels, or images, enabling controlled and contextually relevant outputs instead of random samples. Recent advancements in text-guided diffusion models, such as Glide~\cite{nichol2021glide}, DALL-E 2~\cite{ramesh2022hierarchical}, and Imagen~\cite{saharia2022photorealistic}, have demonstrated exceptional capabilities in generating high-quality, contextually relevant images based on textual descriptions. These models leverage semantic understanding to bridge the gap between language and visual representation, enabling applications in creative design, personalized content creation, and advertising. Image-guided diffusion models, including SR3~\cite{saharia2022image}, Palette~\cite{saharia2022palette}, and Srdiff~\cite{li2022srdiff}, specialize in transforming or enhancing existing images through super-resolution, inpainting, and style transfer. Their ability to generate realistic and detailed outputs makes them valuable for tasks requiring dense image-to-image predictions. The latent diffusion model~\cite{rombach2022high} combines the strengths of both text and image guidance by operating in a lower-dimensional latent space. This approach improves computational efficiency while maintaining high generation quality, making it well-suited for resource-intensive tasks and expanding practical applications in high-quality image synthesis.

\section{Methods}
\subsection{Background}
\vspace{0.1cm}
\paragraph{Diffusion Models}
Starting from a noise-free image $x_0$, the forward diffusion process is a continuously-time stochastic process from time $t = 0$ to $t = 1$, incrementally introducing noise to the image $x_0$ until it becomes pure Gaussian noise at $t = 1$. The distributions of the intermediate noisy images $x(t)$ are given by~\cite{kingma2021variational}:
\begin{equation}
q(x(t)|x_0) = \mathcal{N}(\alpha(t) x_0, \sigma^2(t) \mathbb{I})
\label{eq:forward}
\end{equation}
where $x(0) = x_0$, $\alpha(t)$ and $\sigma(t)$ are differentiable functions, and the signal-to-noise ratio function $\text{SNR}(t) := \frac{\alpha^2(t)}{\sigma^2(t)}$ decreases monotonically from $+\infty$ at $t = 0$ to $0$ at $t = 1$. An equivalent formulation of $x(t)$ is given by the following stochastic differential equation (SDE):
\begin{equation}
dx(t) = f(t) x(t) dt + g(t) dW(t)
\end{equation}
where $f(t) = \frac{\alpha'(t)}{\alpha(t)}$, $g(t) = \sqrt{ (\sigma^2(t))' - 2 \frac{\alpha'(t)}{\alpha(t)} \sigma^2(t)}$, and $W(t)$ is a standard Wiener process from $t = 0$ to $t = 1$. It has been shown that the reverse process of $x(t)$ is given by the following reverse-time SDE~\cite{song2020score}:
\begin{equation}
    dx(t) = [f(t) - g^2(t) \nabla_{x(t)} \log p(x(t)) dt + g(t) d\Bar{W}(t)
\end{equation}
where $x(1) = \mathcal{N}(0, \mathbb{I})$, $d\Bar{W}(t)$ the standard Wiener process backwards from $t = 1$ to $t = 0$, and $ p(x(t))$ denote the probability density of $x(t)$. If the score function $\nabla_{x(t)} \text{log}  p(x(t))$ is known for every $t\in[0,1]$, one can solve the reverse-time SDE to sample new images. 

We can train a model $s_\theta(x(t),t)$ to estimate the score function using the following loss function:

\begin{equation}
    \resizebox{0.8\columnwidth}{!}{
    $\underset{t \in [0,1], x_0 \sim p_{\text{data}}, x(t) \sim q(x(t)|x_0) }
    {\mathbb{E}} \lambda(t)\left\|s_\theta(x(t),t) - \nabla_{x(t)} \log p(x(t)|x_0)\right\|_2^2 $}
\end{equation}

where $p(x(t)|x_0)$ denote the density function of $x(t)$ generated in Eq.~\ref{eq:forward}  and $\lambda(t)$ is a scalar-valued weight function. The training of $s_\theta(x(t),t)$ is straightforward since $q(x(t)|x_0)$ is a Gaussian distribution and its score function has a closed form:
\begin{equation}
\nabla_{x(t)} \text{log} p(x(t)|x_0) = -\frac{x(t) - \alpha(t)x_0}{\sigma^2(t)} = -\frac{\epsilon }{\sigma(t)}
\end{equation}
where $\epsilon$ is a random noise sample from $\mathcal{N}(0, \mathbb{I})$ to generate $x(t)$ from $x_0$. Equivalently, we can use a U-Net denoiser $\epsilon_\theta(x(t),t)$ to estimate the random noise $\epsilon$ (i.e., the score function scaled by $-\sigma(t)$). Following \cite{ho2020denoising}, in this paper, we trained the U-Net denoiser using a simple loss function:
\begin{equation}
\underset{i\in [{1, \cdots, T }], x_0 \sim p_{\text{data}}, \epsilon \sim \mathcal{N}(0, \mathbb{I}) } {\mathbb{E}} [ \epsilon - \epsilon_{\theta}(\alpha(t)x_0 + \sigma(t) \epsilon,t)]^2 
\end{equation}

where $T$ is the total number of time steps and $t = \frac{i}{T}$.

\paragraph{Linear-$\beta$ Noise Scheduler}
\label{sec:noise}
Any noise scheduler, defined by $\alpha(t)$ and $\sigma(t)$), should ensure that $\text{SNR}(1) = \frac{\alpha^2(1)}{\sigma^2(1)} \approx 0$ so that $x(1)$ is pure Gaussian noise. The variance-preserving noise schedulers~\cite{kingma2021variational} meet this requirement, where $\alpha(t)$ is a monotonically decreasing function, $\alpha(0) = 1$, $\alpha(1) = 0$, and $\alpha^2(t) + \sigma^2(t) = 1$ for all $t$. 
This formulation guarantees that the variance of any latent image $x(t)$ is bounded. 
The most popular variance-persevering noise scheduler is the linear-$\beta$ noise scheduler used in the DDPM model~\cite{ho2020denoising}. In this approach, the forward diffusion process was defined as a Markovian process at discrete time steps $i \in \{ 1, 2, \cdots, T\}$ using the following transition kernel:

\begin{equation}
    \resizebox{0.8\columnwidth}{!}{
    $q\left(x\left(\frac{i}{T}\right) \middle| x\left(\frac{i-1}{T}\right)\right) = \mathcal{N}\left(\sqrt{1 - \beta\left(\frac{i}{T}\right)} x\left(\frac{i-1}{T}\right), \beta\left(\frac{i}{T}\right) \mathbb{I}\right)$}
\end{equation}

where $t = \frac{i}{T}$, $\beta(t) = 0.0001 + (0.02 - 0.0001)t$, and the corresponding $\alpha$ noise scheduler $\alpha(t)$ at discrete time steps is given by 
\begin{equation}
    \alpha^{2}\left(\frac{i}{T}\right) = \prod_{j=1}^{i} \left(1 - \beta\left(\frac{j}{T}\right)\right)
    \label{eq:alpha}
\end{equation}
% \begin{equation}
% \alpha(\frac{i}{T}) = \prod_{j = 1}^{i} \big(1 - \beta(\frac{j}{T})\big)
% \label{eq:alpha}
% \end{equation}

To establish feasible noise schedulers using Eq.~\ref{eq:alpha}, the total number of time steps $T$ should be carefully chosen. 
\begin{figure}[!hbt]
\centering
\includegraphics[width=0.85\columnwidth]{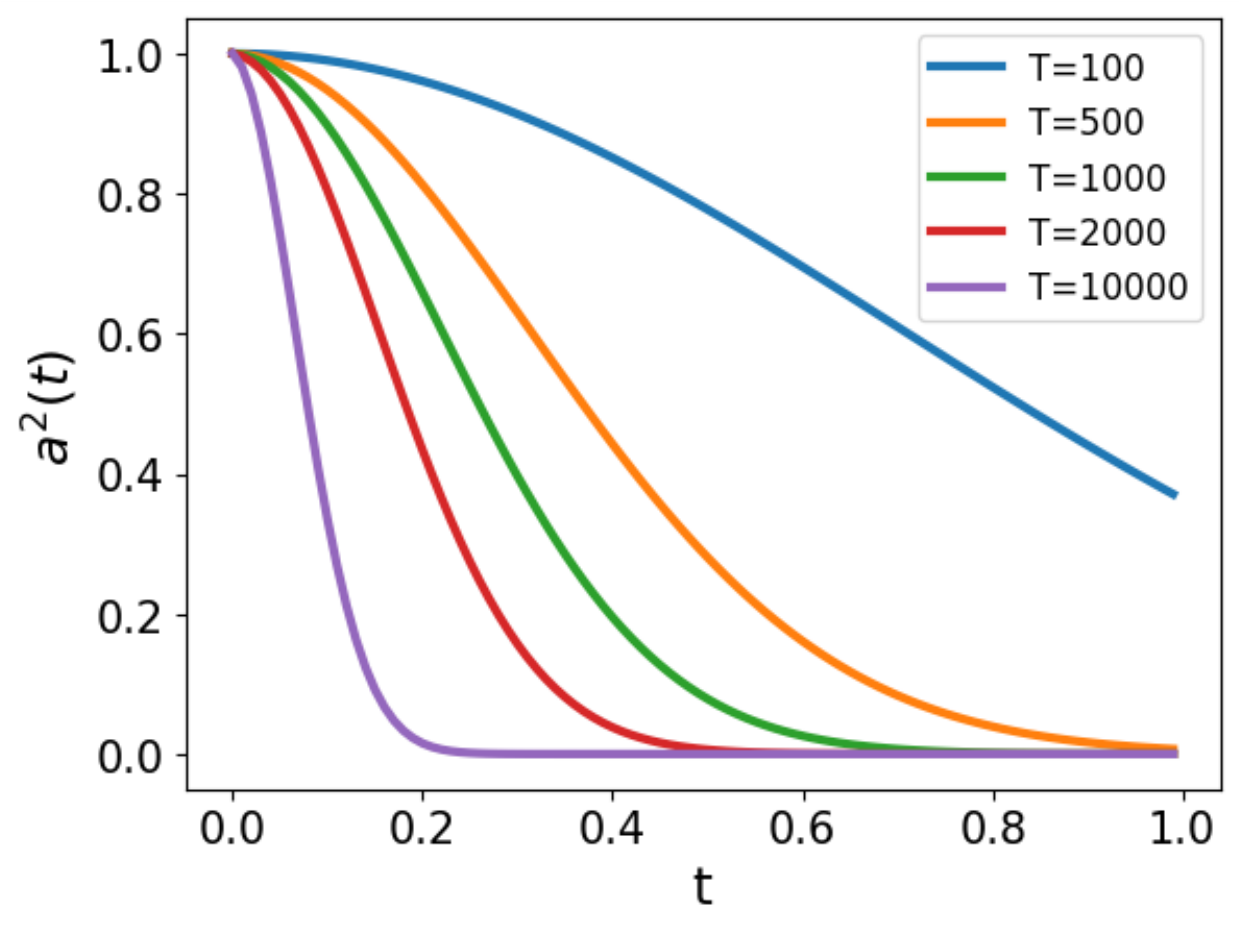}
\caption{Impact of the number of time steps, $T$, on the noise scheduler.}
\label{fig:beta_noise_scheduler}
\end{figure}
Figure~\ref{fig:beta_noise_scheduler} plots $\alpha^2(t)$ for different choices of $T$: 100, 500, 1000, 2000, and 10000.
As we can see, the total number of time steps cannot be too small (e.g., $T = 100$), as this results in the endpoint of the forward diffusion process $x(1)$ not being purely Gaussian. Conversely, if the total number of time steps is too large (e.g., $T = 10000$), $x(t)$ becomes pure Gaussian noise at an early time point (e.g., $t = 0.2$). This makes the training of $\epsilon_\theta(x(t), t)$ at time steps $t > 0.2$ unnecessary since there is no change in the noise level of $x(t)$ beyond this point. The plots in Figure~\ref{fig:beta_noise_scheduler} suggest that a good choice of $T$ lies between 500 and 2000. This explains why a linear-$\beta$ noise scheduler with $T = 1000$ is currently the most popular choice~\cite{ho2020denoising}. In other words, we cannot reduce the number of time steps to 10 using Eq.~\ref{eq:alpha}. In this study, we first define the continuous function $\alpha^2(t)$ and then sample it at discrete time points.

\subsection{Fast Denoising Diffusion Probabilistic Model (Fast-DDPM)}
We have developed the Fast-DDPM model to accelerate the training and sampling of DDPM by utilizing task-specific noise schedulers with only 5-10 time steps. 

\begin{figure}[!htbp]
    \centering
    \begin{subfigure}[b]{0.45\textwidth}
        \centering
        \includegraphics[width=0.8\textwidth]{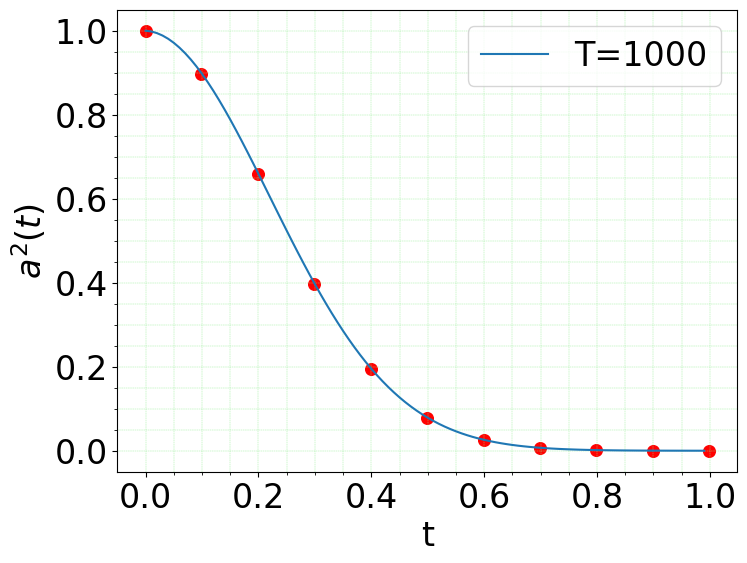}
        \caption{The Uniform Sampling}
        \label{fig:uniform_sampling}
    \end{subfigure}
    \begin{subfigure}[b]{0.45\textwidth}
        \centering
        \includegraphics[width=0.8\textwidth]{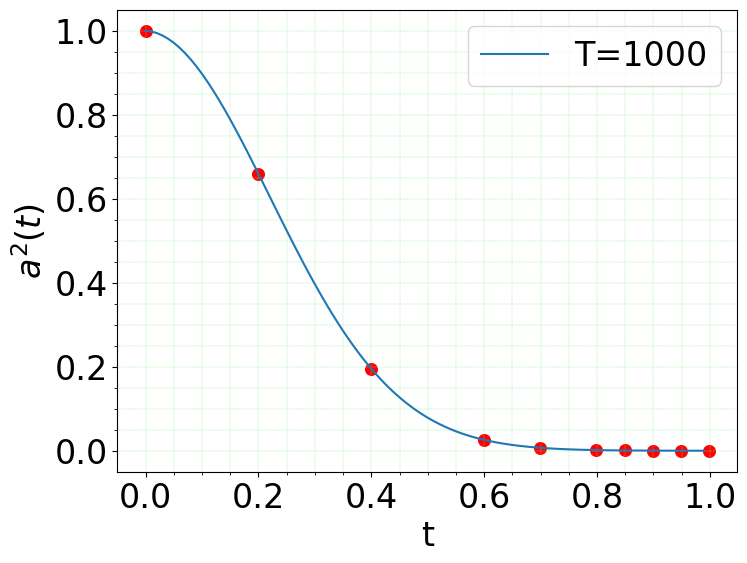}
        \caption{The Non-Uniform Sampling}
        \label{fig:non_uniform_sampling}
    \end{subfigure}
    \caption{Proposed $\alpha$ noise schedulers with uniform and non-uniform sampling between [0,1].}
    \label{fig:different_noise_schedulers}
\end{figure}

\paragraph{Task-Specific $\alpha$ Noise Scheduler}
\label{sec:eff_ns}
Previous studies indicate that images of different sizes require different noise schedulers for optimal results~\cite{chen2023importance, hoogeboom2023simple}. Inspired by this, we propose two $\alpha$ noise schedulers for various medical image-to-image tasks. We first define a smooth, monotonically decreasing function $\alpha^2(t)$ from $t = 0$ to $t = 1$, with boundary conditions $\alpha^2(0) = 1$ and $\alpha^2(1) = 0$. In this paper, we use $\alpha^2(t)$ as defined by the linear-$\beta$ noise scheduler with $T = 1,000$ in Eq.\ref{eq:alpha} (see the green curve in Figure~\ref{fig:beta_noise_scheduler}). Unlike the linear-$\beta$ noise scheduler used in the DDPM model, which uniformly samples 1,000 time points between [0,1], our first proposed noise scheduler uniformly samples 10 time steps between [0,1] (Figure~\ref{fig:different_noise_schedulers}(a)). The second proposed noise scheduler non-uniformly samples 10 time points between [0,1](Figure~\ref{fig:different_noise_schedulers}(b)). 
To optimize efficiency, we designate time step 699 as a boundary, sampling 60\% of the total time steps uniformly from $t>699$ and the remaining 40\% from $t\leq699$. While we do not explicitly define a closed-form formula for the sampling rate as a function of noise level, this heuristic approach follows prior observations~\cite{chen2023importance} that later time steps contribute more to perceptual quality.
Our formulation of noise schedulers satisfies all the desirable properties outlined earlier. Due to the reduced training and sampling time, users have the opportunity to try both noise schedulers for each image-to-image task to determine which one fits their task better.

We provide a rationale for our design. In a related context of diffeomorphic image registration, 10 time points are sufficient to solve an ODE for accurately estimating a complex time-dependent image flow between two images~\cite{miller2006geodesic,shao2021geodesic}. Since we can sample at only 10 discrete time points during inference, we hypothesize that we only need to train the denoiser $\epsilon_{\theta}$ at those 10 time steps, not necessarily at all the 1,000 time points.

\paragraph{Training and Sampling of Fast-DDPM}
In this paper, Fast-DDPM is applied to conditional image-to-image generation. For each image $x_0$ sampled from $p_{\text{data}}(x_0)$, we assume there are one or more associated condition images, denoted by $c$. Here, $c$ may represent either a single image or multiple images, depending on the specific application. We denote the joint distribution of $x_0$ and $c$ as $p_{\text{joint}}(x_0,c)$ and the marginal distribution of the conditional image as $p_c(c)$. We will train a conditional image denoiser $\epsilon_\theta(x(t),c,t)$ that takes the condition $c$ as an additional input to guide the estimation of the score function at each time $t$. We used the following simple loss function that assigns equal weights to different times $t$:
\begin{equation}
\underset{i, (x_0,c), \epsilon} {\mathbb{E}} || \epsilon - \epsilon_{\theta}(\alpha(t)x_0 + \sigma(t) \epsilon,c,t)||^2
\end{equation}
where $T =10$, $i\in \{1,2, \cdots, T\}$, $t = \frac{i}{T}$, $(x_0,c) \sim p_{\text{joint}}(x_0,c)$, and $\epsilon \sim \mathcal{N}(0, \mathbb{I})$. 

The sampling process starts by randomly sampling a Gaussian noise $x(1) \sim \mathcal{N}(0, \mathbb{I}) $ and a condition image $c \sim p_c(c)$. The DDIM sampler \cite{song2020denoising} was used to solve the reverse-time ODE to obtain the noise-free output $x_0$ that corresponds to $c$.
We summarize the training and sampling processes of Fast-DDPM in Algorithm \ref{alg:training} and Algorithm \ref{alg:sampling}, respectively.

\begin{algorithm}[!htb]
    \SetAlgoLined
    \Repeat{converged}{
        $i \in \{1,...,T\}$  \\
        \vspace{0.2\baselineskip} % Decrease vertical space between lines
        $(x_0, c) \sim p_{\text{joint}(x_0,c)}$ \\
        \vspace{0.2\baselineskip} % Decrease vertical space between lines
        $\epsilon \sim \mathcal{N}(0, \mathbb{I})$ \\
        \vspace{0.1\baselineskip} % Decrease vertical space between lines
        $t = \frac{i}{T}$ \\
        \vspace{0.2\baselineskip} % Decrease vertical space between lines
        Gradient descent: $\nabla_{\theta} || \epsilon - \epsilon_{\theta}(\alpha(t)x_0 + \sigma(t) \epsilon,c,t)||^2$
    }
    \caption{Training Fast-DDPM}
    \label{alg:training}
\end{algorithm}

\begin{algorithm}[!htb]
    \SetAlgoLined
    $x(1) \sim \mathcal{N}(0, \mathbb{I})$ \\
    \vspace{-0.05\baselineskip} % Decrease vertical space between lines
    $c \sim p_{\text{c}} (c)$ \\
    \vspace{0.05\baselineskip} % Decrease vertical space between lines
    \For{$i = T, \cdots, 1$}{
    \vspace{-0.1\baselineskip} % Decrease vertical space between lines
        $t = \frac{i}{T}$ \\
        \vspace{-0.1\baselineskip} % Decrease vertical space between lines
        $x(t - \frac{1}{T}) = \frac{\alpha(t - \frac{1}{T})}{\alpha(t)}x(t) + [\sigma(t - \frac{1}{T}) - \frac{\alpha(t - \frac{1}{T})}  {\alpha(t)}\sigma (t) ] \epsilon_{\theta}\big (x(t), c, t)\big )$
    }
    return $x(0)$
    \caption{Sampling Fast-DDPM}
    \label{alg:sampling}
\end{algorithm}

In conditional diffusion-based models, noisy ground truth is used only during training to help the model learn the data distribution across different noise levels. However, during inference, the model does not rely on the noisy ground truth. Instead, it begins from a random Gaussian noise sample and a condition image, refining the noise through iterative steps to generate a clean output. Thus, the use of noisy ground truth in training does not impact real-world applicability.

\section{Experiments}
We evaluated our Fast-DDPM model on three medical image-to-image tasks: multi-image super-resolution, image denoising, and image-to-image translation. 
\subsection{Datasets}
\paragraph{Prostate MRI Dataset for Multi-Image Super-Resolution.}
We used T2-weighted (T2w) prostate MRI scans from the publicly available Prostate-MRI-US-Biopsy dataset~\cite{sonn2013targeted}. The in-plane resolution of each MRI is 0.547mm $\times$ 0.547mm, with a 1.5mm distance between adjacent slices. An image triplet, comprising three consecutive MRI images, serves as a single training or testing example. In this setup, the first and third slices serve as inputs to the model, while the middle slice is considered as the ground truth. 
The goal of this task is to predict the missing information in 3D between any two adjacent slices, thereby enhancing through-plane image resolution. During training, triplets of consecutive MRI slices are extracted, with the first and third slices serving as inputs to the model and the middle slice as the ground truth. A forward diffusion process adds noise to the middle slice, and the denoiser is trained to reconstruct it from the noisy version using the first and third slices as conditions.
We used a total of 6979 image triplets from 120 MRI volumes for training and 580 image triplets from 10 MRI volumes for testing. All MRI slices were resized to $256 \times 256$ and normalized to the range [-1, 1]. 

\paragraph{Low-Dose and Full-Dose Lung CT Dataset for Image Denoising}
We used paired low-dose and normal-dose chest CT image volumes from the publicly available LDCT-and-Projection-data dataset~\cite{mccollough2021low}. The normal-dose CT scans were acquired at routine dose levels, while the low-dose CT scans were reconstructed using simulated lower dose levels, specifically at 10\% of the routine dose.
In this task, the goal is to generate a normal-dose CT scan from its corresponding low-dose scan. The low-dose CT serves as the condition, and the normal-dose scan is used as the ground truth. During training, a noisy version of the ground truth is created via the forward diffusion process and provided to the denoiser as input, with the corresponding low-dose scan serving as the condition.
We randomly selected 38 patients for training and the 10 patients for evaluation, comprising 13,211 and 3,501 2D low-dose and normal-dose CT image pairs, respectively. All images were resized to 256 $\times$ 256 and normalized to [-1, 1].

\paragraph{BraTS Brain MRI Dataset for Image-to-Image Translation}
We used registered T1-weighted (T1w) and T2-weighted (T2w) MR images from the publicly available BraTS 2018 dataset~\cite{menze2014multimodal}. All images were resampled to a resolution of 1 mm³ and skull-stripped. 
In this task, the goal is to convert T1w MRI into T2w MRI, where the T1w MRI serves as the condition and the T2w scan as the ground truth. During training, the denoiser is trained using the noisy ground truth as input, conditioned on the corresponding T1w MRI.
The dataset, obtained from~\cite{kong2021breaking}, comprised 5760 pairs of T1w and T2w MR images for training and 768 pairs for testing. All MRI slices were padded from 224 $\times$ 224 to 256 $\times$ 256 and normalized to  [-1, 1]. 

\subsection{Evaluation Metrics}
In line with prior work in medical image-to-image generation~\cite{armanious2020medgan,kong2021breaking,sood20213d}, we used the peak signal-to-noise ratio (PSNR) and the structural similarity index measure (SSIM) to evaluate the similarity between the generated images and the ground truth images across all three tasks. PSNR measures the ratio between the maximum possible value (power) of the ground truth image and the power of distorting noise:
\begin{equation}
PSNR = 20 \cdot \log_{10} \left( \frac{MAX_{I}}{\sqrt{MSE}} \right)
\end{equation}
where $MAX_{I}$denotes the maximum possible pixel value of the image and MSE denotes the mean squared error. SSIM~\cite{1284395} is a widely used metric that measures the structural similarity between two images and aligns better with human perception of image quality. 
SSIM is defined as:
\begin{equation}
SSIM(x,\hat{x}) = \frac{2\mu_{x}\mu_{\hat{x}}+c_{1}}{\mu^{2}{x}+\mu^{2}{\hat{x}}+c_{1}} \cdot \frac{2\sigma_{x\hat{x}}+c_{2}}{\sigma^{2}{x}+\sigma^{2}{\hat{x}}+c_{2}}
\end{equation}
where $\mu_{x}, \mu_{\hat{x}}$ and $\sigma_{x}, \sigma_{\hat{x}}$ are the means and standard deviations of image $x$ and image $\hat{x}$, respectively; $\sigma_{x\hat{x}}$ denotes the covariance of $x$ and $\hat{x}$. The value of $c_{1}$ is $(k_{1}L)^{2}$ and $c_{2}=(k_{2}L)^{2}$, where $k_1$ is 0.01, $k_2$ is 0.03, and $L$ denotes the largest pixel value of the image $x$.

\subsection{Training Details}
The hyperparameter settings and the U-Net architectures of Fast-DDPM follow those of DDIM~\cite{song2020denoising}. Training was conducted using the Adam optimizer with a learning rate of $2 \times 10^{-4}$. A linear $\beta$ noise scheduler with $T=1000$, $\beta_{1} = 10^{-4}$, and $\beta_{T} = 0.02$, was used to generate the uniform and non-uniform noise schedulers presented in Figure~\ref{fig:different_noise_schedulers}.
For the image super-resolution task, we used the noise scheduler with non-uniform sampling, and for the denoising and translation tasks, we used the noise scheduler with uniform sampling. We used a training batch size of 16 and trained the Fast-DDPM model for 400,000 iterations and the DDPM model for 2 million iterations. All experiments were conducted on a computing node with 4 NVIDIA A100 GPUs, each with 80GB memory. We used Python 3.10.6 and PyTorch 1.12.1 for all experiments.

\section{Results}

For each image-to-image task, we compare Fast-DDPM with a GAN-based method, a CNN-based method, and DDPM~\cite{ho2020denoising} model with ancestral sampling, and two fast sampling methods, DDIM~\cite{song2020denoising} and DPM-Solver++~\cite{lu2022dpm++}. Additionally, to facilitate a more granular analysis, we conduct ablation studies to investigate the impact of the number of time steps and the type of noise scheduler used across all three tasks.

\subsection{Multi-Image Super-Resolution}
Results in Table~\ref{table:sr} show that Fast-DDPM outperformed miSRCNN~\cite{dong2015image} and miSRGAN~\cite{sood20213d}, as well as all three diffusion-based methods: DDPM, DDIM, and DPM-Solver++. Notably, Fast-DDPM significantly improved upon DDPM, increasing PSNR from 25.3 to 27.1 and SSIM from 0.83 to 0.89, while reducing training time from 136 hours to 26 hours and per-volume inference time from 3.7 minutes to 2.3 seconds. Although DPM-Solver++ and DDIM offer similar computational efficiency to Fast-DDPM, their performance in terms of PSNR and SSIM is inferior.

%If we train Fast-DDPM for 2 million iterations (Fast-DDPM*), it maintains the same image quality, demonstrating its efficiency in training.
\begin{table}[!hbt]
\centering
\caption{Comparison of the performance and inference time per image volume (an average of 58 slices) for various multi-image super-resolution methods.}
\label{table:sr}
\setlength{\tabcolsep}{5pt}
\begin{tabular}{ccccc}
\hline
Method                                  & PSNR         & SSIM         & Training time     & Inference Time  \\ \hline
miSRCNN~\cite{dong2015image}            & 26.5              & 0.87              & 1 h               & 0.01s           \\ \hline
miSRGAN~\cite{sood20213d}               & 26.8              & 0.88              & 40 h              & 0.04 s          \\ \hline
DDPM~\cite{ho2020denoising}             & 25.3              & 0.83              & 136 h             & 3.7 m           \\ \hline
DDIM~\cite{song2020denoising}            & 26.5              & 0.88              & 26 h             & 2.3 s           \\ \hline
DPM-Solver++~\cite{lu2022dpm++}             & 26.5              & 0.87              & 26 h             & 2.3 s           \\ \hline
Fast-DDPM                               & \textbf{27.1}     & \textbf{0.89}     & 26 h              & 2.3 s            \\ \hline
%Fast-DDPM*                              & \textbf{27.1}     & \textbf{0.89}     & 139 h             & 2.3 s            \\ \hline
\end{tabular}
\end{table}

\begin{figure}[!hbt]
\centering
\includegraphics[width=0.38\textwidth]{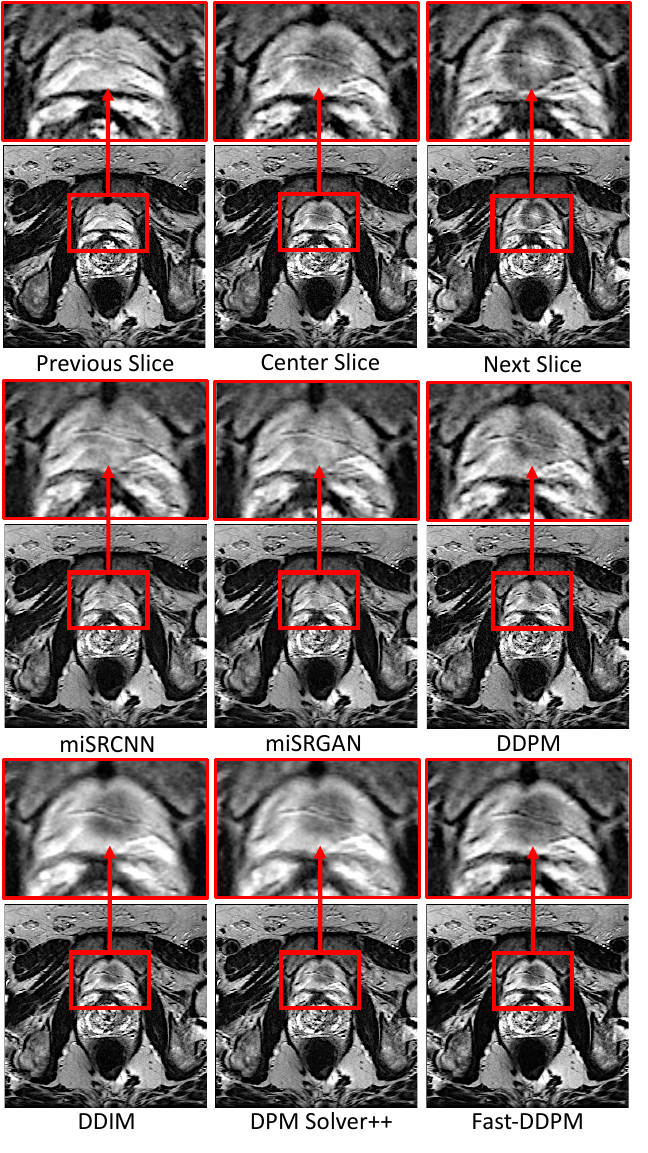}
\caption{Qualitative results of the MR multi-image super-resolution task.} 
\label{fig:sr}
\end{figure}

Figure~\ref{fig:sr} shows the results of various image super-resolution methods on a representative subject. The models received the previous and next slices as inputs, with the center slice serving as the ground truth. It is evident that the miSRCNN and miSRGAN methods failed to accurately reconstruct the shadowed area highlighted by the arrows. In contrast, Fast-DDPM most effectively reconstructed the information missing between the previous and next 2D slices. Interestingly, although DDPM is qualitatively inferior to miSRCNN and miSRGAN, it visually outperformed both methods.
While DDIM and DPM-Solver++ also demonstrated improvements over DDPM, Fast-DDPM consistently exhibited superior visual and quantitative performance.

\subsection{Image Denoising}
The results presented in Table~\ref{tb:denoising} indicate that Fast-DDPM significantly outperformed DDPM, as well as two other prominent CT denoising methods: REDCNN~\cite{chen2017low} and DU-GAN~\cite{huang2021gan}. In terms of computational efficiency, Fast-DDPM reduced inference time by approximately 100-fold and training time by 5-fold compared to DDPM. Furthermore, compared to prior state-of-the-art fast-sampling methods, DPM-Solver++ and DDIM, Fast-DDPM achieved higher PSNR and SSIM scores, setting a new benchmark.

%Notably, the performance of the original DDPM model was inferior to that of REDCNN and DU-GAN.

%Additionally, if the proposed Fast-DDPM model is trained for a longer period (Fast-DDPM*) with the same number of iterations as DDPM, its performance improves slightly, with PSNR increasing from 37.4 to 37.5, while SSIM remains the same.
\begin{table}[!hbt]
\centering
\caption{Comparison of the performance and inference time per image volume (an average of 360 slices) for various CT image denoising methods.}
\label{tb:denoising}
\setlength{\tabcolsep}{5pt}
\begin{tabular}{ccccc}
\hline
Method                                  & PSNR         & SSIM     & Training time & Inference Time  \\ \hline
REDCNN~\cite{chen2017low}               & 36.4              & 0.91          & 3 h           & 0.5 s           \\ \hline
DU-GAN~\cite{huang2021gan}              & 36.3              & 0.90          & 20 h          & 3.8 s           \\ \hline
DDPM~\cite{ho2020denoising}             & 35.4              & 0.87          & 141 h         & 21.4 m           \\ \hline
DDIM~\cite{song2020denoising}            & 37.4              & 0.91          & 26 h         & 12.5 s           \\ \hline
DPM-Solver++~\cite{lu2022dpm++}            & 37.0              & 0.90          & 26 h         & 12.5 s           \\ \hline
Fast-DDPM                               & \textbf{37.5}              & \textbf{0.92} & 26 h          & 12.5 s           \\ \hline
\end{tabular}
\end{table}

\begin{figure}[!hbt]
\centering
\includegraphics[width=0.48\textwidth]{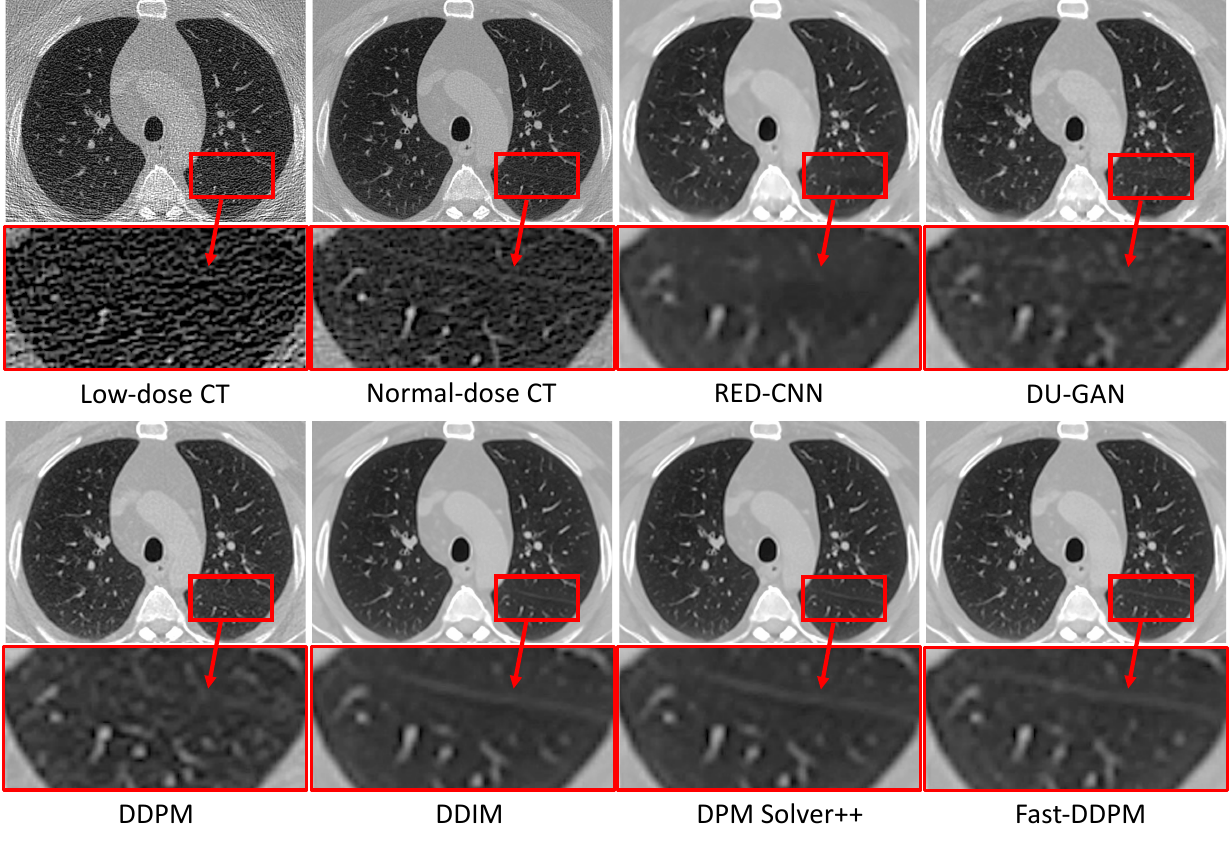}
\caption{Qualitative results of the CT image denoising task.} 
\label{fig:denoising}
\end{figure}

Figure~\ref{fig:denoising} shows the low-dose, normal-dose, and predicted normal-dose CT images generated by various models for a representative subject. The zoomed-in region, highlighted by red rectangular boxes, demonstrates the ability of our Fast-DDPM model to preserve fine structural details in the denoised images compared to other methods. Notably, the lung fissure appears clearest and sharpest in the image produced by Fast-DDPM, while other methods (except for DDIM and DPM-Solver++) exhibit noticeable blurring and loss of detail in this region. A pulmonary fissure is the boundary between lobes of the lung. Preserving fissure details enables automated segmentation of the lung lobes, facilitating more detailed analysis of lung diseases at the lobar level.

\subsection{Image-to-Image Translation}
As shown in Table~\ref{tab:image_translation}, Fast-DDPM outperformed all other leading methods on the image-to-image translation task, including Pix2Pix~\cite{isola2017image}, RegGAN~\cite{kong2021breaking}, DDPM, DDIM, and DPM-Solver++. While DDPM achieved comparable performance, it required a total training time of 135 hours and 22.2 minutes per batch for sampling. In contrast, Fast-DDPM significantly reduced the training time to 27 hours and the sampling time to 13.2 seconds per batch. These improvements highlight the practical advantages of Fast-DDPM in real-world applications.
\begin{table}[!hbt]
\centering
\caption{Comparison of the performance and averaged inference time per batch (360 slices) for various image-to-image translation methods.}
\label{tab:image_translation}
\setlength{\tabcolsep}{5pt}
\begin{tabular}{ccccc}
\hline
Method                                  & PSNR         &  SSIM         & Training time & Inference Time  \\ \hline
Pix2Pix~\cite{isola2017image}           & 25.6              & 0.85              & 6 h           &  3.3 s          \\ \hline
RegGAN~\cite{kong2021breaking}          & 26.0              & 0.86              & 9 h           & 3.1 s            \\ \hline
DDPM~\cite{ho2020denoising}             & \textbf{26.3}     & \textbf{0.89}              & 135 h         & 22.2 m           \\ \hline
DDIM~\cite{song2020denoising}             & 26.2     & 0.88              & 27 h         & 13.2 s           \\ \hline
DPM-Solver++~\cite{lu2022dpm++}             & 26.2     & 0.84              & 27 h         & 13.2 s           \\ \hline
Fast-DDPM                               & \textbf{26.3}     & \textbf{0.89}              & 27 h          & 13.2 s           \\ \hline
\end{tabular}
\end{table}

\begin{figure}[!hbt]
\centering
\includegraphics[width=0.45\textwidth]{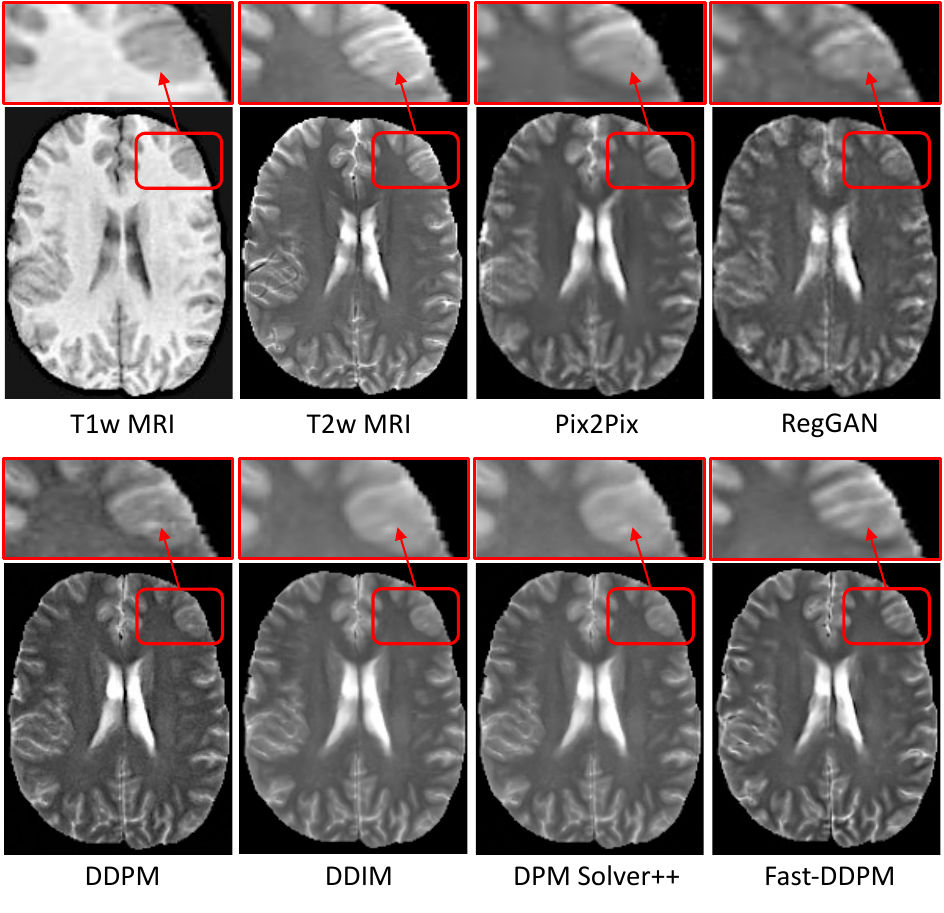}
\caption{Qualitative results of the T1w MRI to T2w MRI translation task.} 
\label{fig:translation}
\end{figure}

Figure~\ref{fig:translation} shows the T2w MR images generated by various image-to-image translation methods for a representative subject. Notably, the predictions from Pix2Pix and RegGAN appear blurry and fail to reconstruct the intricate brain structures highlighted by the arrows. DDIM and DPM-Solver++ also produce blurry predictions and struggle to recover fine details effectively. Although DDPM restores some structural details, the reconstructed T2w MRI exhibits a relatively low signal-to-noise ratio. In contrast, Fast-DDPM produces high-quality translations with the sharpest structural fidelity and minimal noise, demonstrating its superior performance.

\subsection{Ablation Study}
\paragraph{Impact of the Number of Time Steps} We investigated the impact of the number of time steps on the performance of Fast-DDPM. Specifically, we conducted experiments using 3, 5, 7, 10, 20, 50, 100, 200, 500 and 1000 time steps across all three datasets.  
A uniform noise scheduler (Fig. \ref{fig:different_noise_schedulers}a) was used for Denoising and Image-to-Image Translation, while a non-uniform noise scheduler (Fig. \ref{fig:different_noise_schedulers}b) was used for Super-Resolution.

Table~\ref{tab:time_steps} shows that for Super-Resolution and Denoising, model performance improves with the number of time steps up to 10, but declines beyond that. Both tasks refine an input image that already provides a strong structural prior, meaning that excessive diffusion steps may introduce unnecessary noise or blurring, reducing perceptual quality. Since these tasks primarily enhance image fidelity rather than synthesizing entirely new structures, fewer steps enable efficient denoising while preserving fine details.

\begin{table}[!hbt]
\centering
\caption{The impact of the number of time steps on different image-to-image tasks.}
\label{tab:time_steps}
\resizebox{\columnwidth}{!}{%
\begin{tabular}{c|cc|cc|cc}
\hline
\multirow{2}{*}{\begin{tabular}[c]{@{}c@{}}Number of \\ Time Steps\end{tabular}} & \multicolumn{2}{c|}{Super-Resolution} & \multicolumn{2}{c|}{Denoising} & \multicolumn{2}{c}{Translation} \\ \cline{2-7} 
      & PSNR         & SSIM         & PSNR            & SSIM           & PSNR            & SSIM              \\ \hline
3     & 24.8      & 0.79       & 32.7     & 0.83     & 25.3    & 0.86            \\
5     & 26.9      & 0.88       & 37.4     & 0.92     & 25.8            & 0.91           \\
7     & 27.0      & 0.88       & 37.4     & 0.92     & 25.9  & 0.91           \\
10    & \textbf{27.1}         & \textbf{0.89}         &  \textbf{37.5}          & \textbf{0.92}          & 26.3           & 0.91              \\
20    & 26.7      & 0.88       & 37.4            & 0.91           & 26.5            & 0.91                             \\
50    & 26.2      & 0.87       & 36.9            & 0.91           & 26.5            & 0.92                             \\
100   & 25.8      & 0.86       & 36.5            & 0.90           & \textbf{26.6}   & \textbf{0.92}                    \\
200   & 25.6      & 0.85       & 36.0        & 0.89           & 26.6            & 0.91           \\
500     & 25.4     & 0.84         & 35.7    & 0.88           &  26.4    & 0.91           \\
1000 & 25.3 & 0.83 &  35.4 & 0.87 & 26.3 &  0.89\\ \hline
\end{tabular}
}
\end{table}

In contrast, Image-to-Image Translation follows a different trend, achieving peak performance at 100 time steps before declining. Unlike Super-Resolution and Denoising, this task involves generating new image appearances, often requiring more extensive transformations between the condition image and the final output. A higher number of diffusion steps likely allows the model to refine these transformations more effectively, leading to improved synthesis quality. However, beyond 100 steps, additional diffusion may introduce unnecessary noise refinements or reduce structural consistency, leading to a performance drop.

These results highlight that the optimal number of time steps depends on the nature of the task. Tasks focused on enhancing image fidelity (Super-Resolution, Denoising) benefit from fewer steps, whereas tasks requiring structural transformation (Image-to-Image Translation) require more steps to fully reconstruct new details. Future research could explore adaptive strategies to dynamically adjust the number of time steps based on task-specific requirements.

\paragraph{Impact of Noise Scheduler}
Table~\ref{tab:noise_scheduler} illustrates the impact of the two proposed noise schedulers on the performance of Fast-DDPM. The uniform noise scheduler is a better choice for image denoising and image-to-image translation tasks, while the non-uniform noise scheduler performed better in the super-resolution task. These findings highlight the importance of selecting an appropriate noise scheduler tailored to the specific image-to-image task. This adaptability enhances Fast-DDPM’s performance across various medical image processing applications, demonstrating its versatility and effectiveness in handling diverse imaging challenges.

\begin{table}[!hbt]
\centering
\caption{The impact of uniform and non-uniform noise schedulers on different image-to-image tasks.}
\label{tab:noise_scheduler}
\resizebox{\columnwidth}{!}{%
\begin{tabular}{c|cc|cc|cc}
\hline
\multirow{2}{*}{\begin{tabular}[c]{@{}c@{}}Noise \\ Scheduler\end{tabular}} & \multicolumn{2}{c|}{Super-Resolution} & \multicolumn{2}{c|}{Denoising} & \multicolumn{2}{c}{Translation} \\ \cline{2-7} 
                                 & PSNR              & SSIM              & PSNR              & SSIM             & PSNR              & SSIM              \\ \hline
Uniform            & 26.6   & 0.88      & \textbf{37.5}   & \textbf{0.92}      & \textbf{26.3}   & \textbf{0.89}                  \\
Non-uniform        & \textbf{27.1}   & \textbf{0.89}      & 37.3   & 0.91      & 26.1   & 0.88                  \\ \hline
\end{tabular}
}
\end{table}

\section{Discussion}

\subsection{Alignment with Clinical Standards}
In the medical field, image generation and manipulation must adhere to clinical standards and protocols to ensure practical relevance and patient safety. Fast-DDPM addresses this critical need by focusing on clinically relevant tasks such as multi-image super-resolution, image denoising, and image-to-image translation, which directly impact diagnostic accuracy and streamline clinical workflows, thereby enhancing patient care. Unlike general-purpose image generators that prioritize visual diversity and creativity, Fast-DDPM is designed to ensure alignment with ground truth data, evaluated using clinically relevant metrics such as peak signal-to-noise ratio (PSNR) and structural similarity index measure (SSIM). This alignment guarantees that the generated outputs preserve the intensity and structural fidelity essential for clinical decision-making. Furthermore, our approach significantly accelerates both the training and inference of diffusion models, paving the way for real-time clinical applications.

\subsection{Speed-Accuracy Trade-Off}
In unconditional image generation, the model must generate realistic samples entirely from noise, requiring it to learn the full data distribution across all possible variations. This typically necessitates a large number of time steps to incrementally refine details. Reducing the number of steps can disrupt this process, leading to a mismatch between the learned and true distribution and ultimately degrading sample quality. As a result, there is often a trade-off between speed and generation quality.

However, in conditional image generation, the model is guided by an external signal (e.g., a low-resolution version of the target image), significantly reducing the complexity of the learning task. Instead of synthesizing content from scratch, the model learns to refine existing structures. Consequently, using fewer, well-optimized time steps can improve performance by allowing the model to focus its capacity on relevant noise levels rather than being diluted across unnecessary diffusion steps.

We hypothesize that training a diffusion model across a wide range of noise levels without prioritization may introduce excessive redundancy, making it harder to learn distinctive features at different noise levels. Instead, a more selective, step-aware training approach—where priority is given to the noise levels that directly influence the sampling path—can foster more effective representation learning and ultimately lead to superior sample quality.

\subsection{Diffusion Models vs. GANs}
While generative adversarial networks (GANs) have proven effective in various image processing tasks, diffusion models—particularly Fast-DDPM—offer several distinct advantages over GANs. First, Fast-DDPM consistently generates higher-quality samples than GANs, avoiding artifacts that can obscure critical details in medical images. It achieves this without the extended training and sampling times typical of other diffusion models. Second, Fast-DDPM ensures superior training stability, effectively avoiding the mode collapse issue often encountered in GANs due to their adversarial framework. Additionally, Fast-DDPM is more robust to hyperparameter tuning, as its performance does not depend on balancing generator and discriminator losses—a common challenge with GANs. This efficiency, combined with adaptability to applications such as text and audio synthesis, positions Fast-DDPM as a highly promising solution for a wide range of use cases.

\subsection{Limitations and Future Work}
This study has a few limitations that could be addressed in future work. First, the noise schedulers used were manually selected and may not be optimal for the specific tasks. Future work will focus on developing adaptive noise schedulers that can be automatically learned and dynamically adjusted based on the input data and the specific task requirements. Second, this paper focuses on 2D diffusion models, which may not yield optimal performance compared to applying 3D diffusion models. However, model development in 3D requires significant computational resources, which is beyond our current capacity. Nevertheless, the concept of Fast-DDPM can be directly applied to 3D. Future work will focus on generalizing Fast-DDPM to 3D by running the diffusion models in a smaller 3D latent space using autoencoders.

\subsection{Significance and Implications}
The advancements introduced by Fast-DDPM have significant implications for the field of medical imaging. Our results demonstrate that Fast-DDPM achieves state-of-the-art performance in enhancing medical images across tasks such as super-resolution, denoising, and image-to-image translation. These capabilities directly impact clinical workflows by improving image quality, which can enhance diagnostic accuracy, reduce the need for repeat imaging, and support better-informed treatment planning. 
At the same time, Fast-DDPM significantly reduces training and inference time while preserving image quality, addressing a key limitation of traditional diffusion models, where high computational cost and time requirements have limited their practical applications. The dramatic reduction in resource demand makes diffusion models more accessible and practical for real-world clinical applications, particularly in time-sensitive scenarios. The improved efficiency of Fast-DDPM enables real-time image analysis, leading quicker diagnostic decisions and more timely interventions.
Furthermore, Fast-DDPM offers significant flexibility for fine-tuning through adjustments in time steps and noise schedulers. Its design encourages exploration of adaptive noise scheduling methods and task-specific configurations, potentially inspiring future advancements in the field.

\section{Conclusion}
This paper introduces Fast-DDPM, a simple yet effective approach that accelerates both the training and sampling of diffusion models by reducing the number of time steps from 1,000 to 10. Our evaluation on three imaging datasets demonstrates the state-of-the-art performance of Fast-DDPM for medical image-to-image generation tasks. This advancement has the potential to further research in applying diffusion models for real-time medical imaging applications.

\section*{Conflict of interest statement}
The authors have no conflict of interest to declare.

\section*{Acknowledgment}
This work was supported by the Department of Medicine and the Intelligent Clinical Care Center at the University of Florida College of Medicine. The authors express their sincere gratitude to the NVIDIA AI Technology Center at the University of Florida for their invaluable feedback, technical guidance, and support throughout this project.

\bibliographystyle{plain}
\bibliography{refs}

\begin{thebibliography}{10}

\bibitem{armanious2020medgan}
Karim Armanious, Chenming Jiang, Marc Fischer, Thomas K{\"u}stner, Tobias Hepp, Konstantin Nikolaou, Sergios Gatidis, and Bin Yang.
\newblock Medgan: Medical image translation using gans.
\newblock {\em Computerized medical imaging and graphics}, 79:101684, 2020.

\bibitem{10385176}
Chentao Cao, Zhuo-Xu Cui, Yue Wang, Shaonan Liu, Taijin Chen, Hairong Zheng, Dong Liang, and Yanjie Zhu.
\newblock High-frequency space diffusion model for accelerated mri.
\newblock {\em IEEE Transactions on Medical Imaging}, pages 1--1, 2024.

\bibitem{chen2017low}
Hu~Chen, Yi~Zhang, Mannudeep~K Kalra, Feng Lin, Yang Chen, Peixi Liao, Jiliu Zhou, and Ge~Wang.
\newblock Low-dose ct with a residual encoder-decoder convolutional neural network.
\newblock {\em IEEE transactions on medical imaging}, 36(12):2524--2535, 2017.

\bibitem{chen2023importance}
Ting Chen.
\newblock On the importance of noise scheduling for diffusion models.
\newblock {\em arXiv preprint arXiv:2301.10972}, 2023.

\bibitem{9941138}
Hyungjin Chung, Eun~Sun Lee, and Jong~Chul Ye.
\newblock Mr image denoising and super-resolution using regularized reverse diffusion.
\newblock {\em IEEE Transactions on Medical Imaging}, 42(4):922--934, 2023.

\bibitem{dhariwal2021diffusion}
Prafulla Dhariwal and Alexander Nichol.
\newblock Diffusion models beat gans on image synthesis.
\newblock {\em Advances in neural information processing systems}, 34:8780--8794, 2021.

\bibitem{dong2015image}
Chao Dong, Chen~Change Loy, Kaiming He, and Xiaoou Tang.
\newblock Image super-resolution using deep convolutional networks.
\newblock {\em IEEE transactions on pattern analysis and machine intelligence}, 38(2):295--307, 2015.

\bibitem{gong2023pet}
Kuang Gong, Keith Johnson, Georges El~Fakhri, Quanzheng Li, and Tinsu Pan.
\newblock Pet image denoising based on denoising diffusion probabilistic model.
\newblock {\em European Journal of Nuclear Medicine and Molecular Imaging}, pages 1--11, 2023.

\bibitem{GUNGOR2023102872}
Alper Güngör, Salman~UH Dar, Şaban Öztürk, Yilmaz Korkmaz, Hasan~A. Bedel, Gokberk Elmas, Muzaffer Ozbey, and Tolga Çukur.
\newblock Adaptive diffusion priors for accelerated mri reconstruction.
\newblock {\em Medical Image Analysis}, 88:102872, 2023.

\bibitem{ho2020denoising}
Jonathan Ho, Ajay Jain, and Pieter Abbeel.
\newblock Denoising diffusion probabilistic models.
\newblock {\em Advances in neural information processing systems}, 33:6840--6851, 2020.

\bibitem{hoogeboom2023simple}
Emiel Hoogeboom, Jonathan Heek, and Tim Salimans.
\newblock simple diffusion: End-to-end diffusion for high resolution images.
\newblock In {\em International Conference on Machine Learning}, pages 13213--13232. PMLR, 2023.

\bibitem{hu2023conditional}
Xinrong Hu, Yu-Jen Chen, Tsung-Yi Ho, and Yiyu Shi.
\newblock Conditional diffusion models for weakly supervised medical image segmentation, 2023.

\bibitem{huang2021gan}
Zhizhong Huang, Junping Zhang, Yi~Zhang, and Hongming Shan.
\newblock Du-gan: Generative adversarial networks with dual-domain u-net-based discriminators for low-dose ct denoising.
\newblock {\em IEEE Transactions on Instrumentation and Measurement}, 71:1--12, 2021.

\bibitem{10.1007/978-3-031-45673-2_37}
Hasan Iqbal, Umar Khalid, Chen Chen, and Jing Hua.
\newblock Unsupervised anomaly detection in medical images using masked diffusion model.
\newblock In Xiaohuan Cao, Xuanang Xu, Islem Rekik, Zhiming Cui, and Xi~Ouyang, editors, {\em Machine Learning in Medical Imaging}, pages 372--381, Cham, 2024. Springer Nature Switzerland.

\bibitem{isola2017image}
Phillip Isola, Jun-Yan Zhu, Tinghui Zhou, and Alexei~A Efros.
\newblock Image-to-image translation with conditional adversarial networks.
\newblock In {\em Proceedings of the IEEE conference on computer vision and pattern recognition}, pages 1125--1134, 2017.

\bibitem{10.1007/978-3-031-19821-2_20}
Boah Kim, Inhwa Han, and Jong~Chul Ye.
\newblock Diffusemorph: Unsupervised deformable image registration using diffusion model.
\newblock In Shai Avidan, Gabriel Brostow, Moustapha Ciss{\'e}, Giovanni~Maria Farinella, and Tal Hassner, editors, {\em Computer Vision -- ECCV 2022}, pages 347--364, Cham, 2022. Springer Nature Switzerland.

\bibitem{kim2023diffusion}
Boah Kim, Yujin Oh, and Jong~Chul Ye.
\newblock Diffusion adversarial representation learning for self-supervised vessel segmentation.
\newblock In {\em The Eleventh International Conference on Learning Representations}, 2023.

\bibitem{Kim_2024_WACV}
Jonghun Kim and Hyunjin Park.
\newblock Adaptive latent diffusion model for 3d medical image to image translation: Multi-modal magnetic resonance imaging study.
\newblock In {\em Proceedings of the IEEE/CVF Winter Conference on Applications of Computer Vision (WACV)}, pages 7604--7613, January 2024.

\bibitem{kingma2021variational}
Diederik Kingma, Tim Salimans, Ben Poole, and Jonathan Ho.
\newblock Variational diffusion models.
\newblock {\em Advances in neural information processing systems}, 34:21696--21707, 2021.

\bibitem{kong2021breaking}
Lingke Kong, Chenyu Lian, Detian Huang, ZhenJiang Li, Yanle Hu, and Qichao Zhou.
\newblock Breaking the dilemma of medical image-to-image translation.
\newblock In {\em Thirty-Fifth Conference on Neural Information Processing Systems}, 2021.

\bibitem{kong2021fast}
Zhifeng Kong and Wei Ping.
\newblock On fast sampling of diffusion probabilistic models, 2021.

\bibitem{li2022srdiff}
Haoying Li, Yifan Yang, Meng Chang, Shiqi Chen, Huajun Feng, Zhihai Xu, Qi~Li, and Yueting Chen.
\newblock Srdiff: Single image super-resolution with diffusion probabilistic models.
\newblock {\em Neurocomputing}, 479:47--59, 2022.

\bibitem{Q-Diffusion_2023_ICCV}
Xiuyu Li, Yijiang Liu, Long Lian, Huanrui Yang, Zhen Dong, Daniel Kang, Shanghang Zhang, and Kurt Keutzer.
\newblock Q-diffusion: Quantizing diffusion models.
\newblock In {\em Proceedings of the IEEE/CVF International Conference on Computer Vision (ICCV)}, pages 17535--17545, October 2023.

\bibitem{Liu_Reconstruction_2023_ICCV}
Jiaming Liu, Rushil Anirudh, Jayaraman~J. Thiagarajan, Stewart He, K~Aditya Mohan, Ulugbek~S. Kamilov, and Hyojin Kim.
\newblock Dolce: A model-based probabilistic diffusion framework for limited-angle ct reconstruction.
\newblock In {\em Proceedings of the IEEE/CVF International Conference on Computer Vision (ICCV)}, pages 10498--10508, October 2023.

\bibitem{liu2022flow}
Xingchao Liu, Chengyue Gong, and Qiang Liu.
\newblock Flow straight and fast: Learning to generate and transfer data with rectified flow, 2022.

\bibitem{lu2022dpm}
Cheng Lu, Yuhao Zhou, Fan Bao, Jianfei Chen, Chongxuan Li, and Jun Zhu.
\newblock Dpm-solver: A fast ode solver for diffusion probabilistic model sampling in around 10 steps.
\newblock {\em Advances in Neural Information Processing Systems}, 35:5775--5787, 2022.

\bibitem{lu2022dpm++}
Cheng Lu, Yuhao Zhou, Fan Bao, Jianfei Chen, Chongxuan Li, and Jun Zhu.
\newblock Dpm-solver++: Fast solver for guided sampling of diffusion probabilistic models.
\newblock {\em arXiv preprint arXiv:2211.01095}, 2022.

\bibitem{lyu2022conversion}
Qing Lyu and Ge~Wang.
\newblock Conversion between ct and mri images using diffusion and score-matching models, 2022.

\bibitem{10.1007/978-3-031-43898-1_1}
Jun Ma, Yuanzhi Zhu, Chenyu You, and Bo~Wang.
\newblock Pre-trained diffusion models for plug-and-play medical image enhancement.
\newblock In Hayit Greenspan, Anant Madabhushi, Parvin Mousavi, Septimiu Salcudean, James Duncan, Tanveer Syeda-Mahmood, and Russell Taylor, editors, {\em Medical Image Computing and Computer Assisted Intervention -- MICCAI 2023}, pages 3--13, Cham, 2023. Springer Nature Switzerland.

\bibitem{ma2024deepcache}
Xinyin Ma, Gongfan Fang, and Xinchao Wang.
\newblock Deepcache: Accelerating diffusion models for free.
\newblock In {\em Proceedings of the IEEE/CVF Conference on Computer Vision and Pattern Recognition}, pages 15762--15772, 2024.

\bibitem{mccollough2021low}
C~McCollough, B~Chen, D~Holmes, X~Duan, Z~Yu, L~Xu, S~Leng, and J~Fletcher.
\newblock Low dose ct image and projection data (ldct-and-projection-data)(version 4).
\newblock {\em Med. Phys}, 48:902--911, 2021.

\bibitem{meng2023distillation}
Chenlin Meng, Robin Rombach, Ruiqi Gao, Diederik Kingma, Stefano Ermon, Jonathan Ho, and Tim Salimans.
\newblock On distillation of guided diffusion models.
\newblock In {\em Proceedings of the IEEE/CVF Conference on Computer Vision and Pattern Recognition}, pages 14297--14306, 2023.

\bibitem{menze2014multimodal}
Bjoern~H Menze, Andras Jakab, Stefan Bauer, Jayashree Kalpathy-Cramer, Keyvan Farahani, Justin Kirby, Yuliya Burren, Nicole Porz, Johannes Slotboom, Roland Wiest, et~al.
\newblock The multimodal brain tumor image segmentation benchmark (brats).
\newblock {\em IEEE transactions on medical imaging}, 34(10):1993--2024, 2014.

\bibitem{miller2006geodesic}
Michael~I Miller, Alain Trouv{\'e}, and Laurent Younes.
\newblock Geodesic shooting for computational anatomy.
\newblock {\em Journal of mathematical imaging and vision}, 24:209--228, 2006.

\bibitem{nichol2021glide}
Alex Nichol, Prafulla Dhariwal, Aditya Ramesh, Pranav Shyam, Pamela Mishkin, Bob McGrew, Ilya Sutskever, and Mark Chen.
\newblock Glide: Towards photorealistic image generation and editing with text-guided diffusion models.
\newblock {\em arXiv preprint arXiv:2112.10741}, 2021.

\bibitem{nichol2021improved}
Alexander~Quinn Nichol and Prafulla Dhariwal.
\newblock Improved denoising diffusion probabilistic models.
\newblock In {\em International conference on machine learning}, pages 8162--8171. PMLR, 2021.

\bibitem{ozbey2023unsupervised}
Muzaffer {\"O}zbey, Onat Dalmaz, Salman~UH Dar, Hasan~A Bedel, {\c{S}}aban {\"O}zturk, Alper G{\"u}ng{\"o}r, and Tolga {\c{C}}ukur.
\newblock Unsupervised medical image translation with adversarial diffusion models.
\newblock {\em IEEE Transactions on Medical Imaging}, 2023.

\bibitem{ramesh2022hierarchical}
Aditya Ramesh, Prafulla Dhariwal, Alex Nichol, Casey Chu, and Mark Chen.
\newblock Hierarchical text-conditional image generation with clip latents.
\newblock {\em arXiv preprint arXiv:2204.06125}, 1(2):3, 2022.

\bibitem{rombach2022high}
Robin Rombach, Andreas Blattmann, Dominik Lorenz, Patrick Esser, and Bj{\"o}rn Ommer.
\newblock High-resolution image synthesis with latent diffusion models.
\newblock In {\em Proceedings of the IEEE/CVF conference on computer vision and pattern recognition}, pages 10684--10695, 2022.

\bibitem{saharia2022palette}
Chitwan Saharia, William Chan, Huiwen Chang, Chris Lee, Jonathan Ho, Tim Salimans, David Fleet, and Mohammad Norouzi.
\newblock Palette: Image-to-image diffusion models.
\newblock In {\em ACM SIGGRAPH 2022 conference proceedings}, pages 1--10, 2022.

\bibitem{saharia2022photorealistic}
Chitwan Saharia, William Chan, Saurabh Saxena, Lala Li, Jay Whang, Emily~L Denton, Kamyar Ghasemipour, Raphael Gontijo~Lopes, Burcu Karagol~Ayan, Tim Salimans, et~al.
\newblock Photorealistic text-to-image diffusion models with deep language understanding.
\newblock {\em Advances in neural information processing systems}, 35:36479--36494, 2022.

\bibitem{saharia2022image}
Chitwan Saharia, Jonathan Ho, William Chan, Tim Salimans, David~J Fleet, and Mohammad Norouzi.
\newblock Image super-resolution via iterative refinement.
\newblock {\em IEEE Transactions on Pattern Analysis and Machine Intelligence}, 45(4):4713--4726, 2022.

\bibitem{salimans2022progressive}
Tim Salimans and Jonathan Ho.
\newblock Progressive distillation for fast sampling of diffusion models.
\newblock {\em arXiv preprint arXiv:2202.00512}, 2022.

\bibitem{sauer2025adversarial}
Axel Sauer, Dominik Lorenz, Andreas Blattmann, and Robin Rombach.
\newblock Adversarial diffusion distillation.
\newblock In {\em European Conference on Computer Vision}, pages 87--103. Springer, 2025.

\bibitem{Shang_2023_CVPR}
Yuzhang Shang, Zhihang Yuan, Bin Xie, Bingzhe Wu, and Yan Yan.
\newblock Post-training quantization on diffusion models.
\newblock In {\em Proceedings of the IEEE/CVF Conference on Computer Vision and Pattern Recognition (CVPR)}, pages 1972--1981, June 2023.

\bibitem{shao2021geodesic}
Wei Shao, Yue Pan, Oguz~C Durumeric, Joseph~M Reinhardt, John~E Bayouth, Mirabela Rusu, and Gary~E Christensen.
\newblock Geodesic density regression for correcting 4dct pulmonary respiratory motion artifacts.
\newblock {\em Medical image analysis}, 72:102140, 2021.

\bibitem{song2020denoising}
Jiaming Song, Chenlin Meng, and Stefano Ermon.
\newblock Denoising diffusion implicit models.
\newblock {\em arXiv preprint arXiv:2010.02502}, 2020.

\bibitem{song2023consistency}
Yang Song, Prafulla Dhariwal, Mark Chen, and Ilya Sutskever.
\newblock Consistency models.
\newblock {\em arXiv preprint arXiv:2303.01469}, 2023.

\bibitem{song2020score}
Yang Song, Jascha Sohl-Dickstein, Diederik~P Kingma, Abhishek Kumar, Stefano Ermon, and Ben Poole.
\newblock Score-based generative modeling through stochastic differential equations.
\newblock {\em arXiv preprint arXiv:2011.13456}, 2020.

\bibitem{sonn2013targeted}
Geoffrey~A Sonn, Shyam Natarajan, Daniel~JA Margolis, Malu MacAiran, Patricia Lieu, Jiaoti Huang, Frederick~J Dorey, and Leonard~S Marks.
\newblock Targeted biopsy in the detection of prostate cancer using an office based magnetic resonance ultrasound fusion device.
\newblock {\em The Journal of urology}, 189(1):86--92, 2013.

\bibitem{sood20213d}
Rewa~R Sood, Wei Shao, Christian Kunder, Nikola~C Teslovich, Jeffrey~B Wang, Simon~JC Soerensen, Nikhil Madhuripan, Anugayathri Jawahar, James~D Brooks, Pejman Ghanouni, et~al.
\newblock 3d registration of pre-surgical prostate mri and histopathology images via super-resolution volume reconstruction.
\newblock {\em Medical image analysis}, 69:101957, 2021.

\bibitem{10.1007/978-3-031-43999-5_42}
Jueqi Wang, Jacob Levman, Walter Hugo~Lopez Pinaya, Petru-Daniel Tudosiu, M.~Jorge Cardoso, and Razvan Marinescu.
\newblock Inversesr: 3d brain mri super-resolution using a latent diffusion model.
\newblock In Hayit Greenspan, Anant Madabhushi, Parvin Mousavi, Septimiu Salcudean, James Duncan, Tanveer Syeda-Mahmood, and Russell Taylor, editors, {\em Medical Image Computing and Computer Assisted Intervention -- MICCAI 2023}, pages 438--447, Cham, 2023. Springer Nature Switzerland.

\bibitem{1284395}
Zhou Wang, A.C. Bovik, H.R. Sheikh, and E.P. Simoncelli.
\newblock Image quality assessment: from error visibility to structural similarity.
\newblock {\em IEEE Transactions on Image Processing}, 13(4):600--612, 2004.

\bibitem{watson2022learning}
Daniel Watson, William Chan, Jonathan Ho, and Mohammad Norouzi.
\newblock Learning fast samplers for diffusion models by differentiating through sample quality.
\newblock In {\em International Conference on Learning Representations}, 2022.

\bibitem{10.1007/978-3-031-16452-1_4}
Julia Wolleb, Florentin Bieder, Robin Sandk{\"u}hler, and Philippe~C. Cattin.
\newblock Diffusion models for medical anomaly detection.
\newblock In Linwei Wang, Qi~Dou, P.~Thomas Fletcher, Stefanie Speidel, and Shuo Li, editors, {\em Medical Image Computing and Computer Assisted Intervention -- MICCAI 2022}, pages 35--45, Cham, 2022. Springer Nature Switzerland.

\bibitem{WU2023104901}
Zhanxiong Wu, Xuanheng Chen, Sangma Xie, Jian Shen, and Yu~Zeng.
\newblock Super-resolution of brain mri images based on denoising diffusion probabilistic model.
\newblock {\em Biomedical Signal Processing and Control}, 85:104901, 2023.

\bibitem{xing2023diffunet}
Zhaohu Xing, Liang Wan, Huazhu Fu, Guang Yang, and Lei Zhu.
\newblock Diff-unet: A diffusion embedded network for volumetric segmentation, 2023.

\bibitem{xu2024maediff}
Rui Xu, Yunke Wang, and Bo~Du.
\newblock Maediff: Masked autoencoder-enhanced diffusion models for unsupervised anomaly detection in brain images, 2024.

\bibitem{zheng2023fast}
Hongkai Zheng, Weili Nie, Arash Vahdat, Kamyar Azizzadenesheli, and Anima Anandkumar.
\newblock Fast sampling of diffusion models via operator learning.
\newblock In {\em International Conference on Machine Learning}, pages 42390--42402. PMLR, 2023.

\end{thebibliography}

\end{document}